\documentclass[useAMS, usenatbib]{mn2e}

\usepackage[dvips]{graphicx,color}
\usepackage[]{txfonts}

\def\gad{\hat{\gamma}}
\newcommand{\dsfrac}[2]{\displaystyle{\frac{#1}{#2}}}
\newcommand{\eref}[1]{(\ref{#1})}
\newcommand{\secref}[1]{Section~\ref{#1}}
\newcommand{\figref}[1]{Fig.~\ref{#1}}

\title[Efficiency of internal shocks]{On the dynamic efficiency of internal shocks in magnetized relativistic outflows}
\author[P. Mimica and M. A. Aloy]{P. Mimica$^{1}$\thanks{E-mail: Petar.Mimica@uv.es} and M. A. Aloy$^{1}$\\
$^{1}$Departamento de Astronom\'ia y Astrof\'isica, Universidad de Valencia, 46100, Burjassot, Spain}

\begin{document}

\maketitle

\label{firstpage}

\begin{abstract}
  We study the dynamic efficiency of conversion of
  kinetic-to-thermal/magnetic energy of internal shocks in
  relativistic magnetized outflows. We model internal shocks as being
  caused by collisions of shells of plasma with the same energy flux
  and a non-zero relative velocity. The contact surface, where the
  interaction between the shells takes place, can break up either into
  two oppositely moving shocks (in the frame where the contact surface
  is at rest), or into a reverse shock and a forward rarefaction. We
  find that for moderately magnetized shocks (magnetization
  $\sigma\simeq 0.1$), the dynamic efficiency in a single two-shell
  interaction can be as large as $40\%$. Thus, the dynamic efficiency
  of moderately magnetized shocks is larger than in the corresponding
  unmagnetized two-shell interaction. If the slower shell propagates
  with a sufficiently large velocity, the efficiency is only weakly
  dependent on its Lorentz factor. Consequently, the dynamic
  efficiency of shell interactions in the magnetized flow of blazars
  and gamma-ray bursts is effectively the same. These results are
    quantitatively rather independent on the equation of state of the
    plasma. The radiative efficiency of the process is expected to be
  a fraction $f_r<1$ of the estimated dynamic one, the exact value of
  $f_r$ depending on the particularities of the emission processes
  which radiate away the thermal or magnetic energy of the shocked
  states.
\end{abstract}


\section{Introduction}

Internal shocks \citep{Rees:1994ca} are invoked to explain the
variability of blazars \citep[see
e.g.,][]{Spada:2001p815,Mimica:2005sp} and the light curves of the
prompt phase of gamma-ray bursts (GRBs)
\citep{Sari:1995oq,Sari:1997p766,Daigne:1998wq}. A possible problem in
this model is the question whether this mechanism is efficient enough
to explain the relation between the observed energies both in the
prompt GRB phase and in the afterglow (see e.g.,
\citealt{Kobayashi:1997p657} (KPS97), \citealt{Beloborodov:2000p632},
\citealt{Kobayashi:2000p599}, \citealt{Fan:2006p1375}). To asses the
efficiency of the internal shock model, most of the previous works
focus on the comparison between the observed light curves and the
model predictions employing a simple inelastic collision of two-point
masses (KPS97; \citealp{Lazzati:1999p1360}; \citealt{Nakar:2002p1323};
\citealt{Tanihata:2003p1291}; \citealt{Zhang:2004p1381}). Less
attention has been paid to the hydrodynamic effects during the shell
collision \citep[but
see][]{Kobayashi:2000p599,Kino:2004p811,Mimica:2004fy,Mimica:2005sp,Bosnjak:2009p1427}.

The ejecta in GRBs and blazars may be rather magnetized, particularly
if they are originated as a Poynting-flux-dominated flow
\citep[e.g.,][]{Usov:1992hp} Forming shocks in highly magnetized media
is challenging \citep{Rees:1974yq,Kennel:1984kx}. Therefore, to
account for the observed phenomenology it is necessary to address how
efficient the process of internal collisions in arbitrarily magnetized
flows is. This question has been partly considered by a few recent
works \citep[e.g.,][]{Fan:2004p1007,Mimica:2007db}.

The base to study the efficiency of internal collisions is the
determination of the dynamic efficiency of a single binary collision,
i.e., the efficiency of converting the kinetic energy of the colliding
fluid into thermal and/or magnetic energy. Note that the radiative
efficiency (i.e., the efficiency of converting the kinetic energy of
the flow into radiation) is expected to be somewhat smaller. According
to, e.g., \citet{Panaitescu:1999p1022} and \citet{Kumar:1999p1084}, it
can be as low as $f_r\sim 0.1$. As we shall show in this paper,
  binary collisions in relativistic, magnetized flows can be an efficient
  enough way to dissipate a major fraction of the bulk kinetic energy
  of a relativistic ejecta. Therefore, it will depend on the
  efficiency of the particular radiation mechanism, that produces the
  observed emission (i.e., on the factor $f_r$), that the model of
  internal shocks be efficient enough to explain the observations
  (particularly, the distribution of energies between the prompt GRB
  phase and the afterglow phase).

We model internal shocks as shells of plasma with the same energy flux
and a non-zero relative velocity. The contact surface, where the
interaction between the shells takes place, can break up either into
two oppositely moving shocks (in the frame where the contact surface
is at rest), or into a reverse shock and a forward rarefaction. The
determination of whether one or the other possibility occurs is
computed by estimating the invariant relative velocity between the
fastest and the slowest shell, i.e., by solving the Riemann problem
posed by the piecewise uniform states given by the physical quantities
on the two interacting shells (\secref{sec:riemann}). In
\secref{sec:edissipation} we define precisely the notion of dynamic
efficiency, both for shocks and rarefactions.  We perform a parametric
study of the binary shell collision dynamic efficiency in
\secref{sec:parametric}. The discussion and conclusions are listed in
\secref{sec:discussion}. Finally, we have extended our analysis
  on the dynamic efficiency of internal shocks in magnetized,
  relativistic plasma to consider more realistic equations of state in
  the Appendix.

\section{Relativistic magnetohydrodynamic Riemann problem}
\label{sec:riemann}

We model the interaction between parts of the outflow with varying
properties by considering Riemann problems, i.e. relativistic
magnetohydrodynamic initial-value problems with two constant states
separated by a discontinuity in planar symmetry. We note, that we
  could use a more sophisticated approach consisting on performing
  numerical relativistic magnetohydrodynamic (RMHD) simulations of the
  interaction of parts of the outflow with different
  velocities. However, such an approach demands huge computational
  resources (even performing one dimensional simulations using the
  same code as in \citealp{Mimica:2009qa}), and we are interested in
  sampling very finely a large parameter space with our models. Apart
  from this numerical reason, it is in order to point out that, by the
  internal shock phase, the lateral expansion of the flow is very
  small, since the flow is probably cold and ultrarelativistic. Thus,
  a description of the interactions assuming planar symmetry suffices
  to compute the dynamic efficiency of such interactions (rather than
  a more complex spherically symmetric approach).

In the following we use subscripts $L$
and $R$ to denote properties of the (faster) left and (slower) right
state, respectively. To avoid repeated writing of a factor $4\pi$ and
the speed of light $c$, we normalize the rest-mass density $\rho$ to
$\rho_R$, the energy density to $\rho_R c^2$ and the magnetic field
strength to $c\sqrt{4\pi \rho_R}$.

\subsection{Initial states of the Riemann problem}
\label{sec:initial}

For the initial thermal pressure of both states we assume that it is
small fraction of the density, $p_L = \chi \rho_L$ and $p_R =
\chi$. We assume magnetic fields perpendicular to the direction of the
flow propagation.  The remaining parameters determining the  RMHD
Riemann problem are: the density contrast $\rho_L$, the Lorentz factor
of the right state $\Gamma_R$, the relative Lorentz factor difference
$\Delta g := (\Gamma_L - \Gamma_R)/\Gamma_R$, and the magnetizations
of left and right states, $\sigma_L := B_L^2/(\Gamma_R^2(1+\Delta
g)^2\rho_L)$ and $\sigma_R := B_R^2/\Gamma_R^2$, where $B_L$ and $B_R$
are the lab frame magnetic field strengths of left and right states,
respectively. Furthermore, we define the total (thermal + magnetic)
pressure
\begin{equation}\label{eq:pstar}
p^* := p + \dsfrac{B^2}{2\Gamma^2} = p + \dsfrac{\sigma\rho}{2}\, ,
\end{equation}
the total specific enthalpy 
%(assuming an ideal gas equation of state
%with  constant adiabatic index $\gad$ and defining $\gad':=\gad /(\gad - 1)$)
%
\begin{equation}\label{eq:hstar}
h^*:=  1 + \epsilon + p / \rho + \sigma \, ,
\end{equation}
and 
\begin{equation}\label{eq:estar}
e^*:= \rho (1 + \epsilon) + \dsfrac{\sigma\rho}{2}\, .
\end{equation}
where $\epsilon$ denotes the specific internal energy and is dependant on the equation of state used (see \secref{sec:outflow}).

The general solution of a RMHD Riemann problem was found by
  \citet{Giacomazzo:2006p2513}, and recently used in RMHD numerical
  codes by e.g., \citet{vanderHolst:2008p2553}. However, here we deal
  with a degenerate RMHD configuration, which solution was first
found by \citet{Romero:2005zr}. The typical structure of the flow
  after the break up of the initial discontinuity consists of two
initial states, and two intermediate states separated by a contact
discontinuity (CD). The total pressure and velocity are the same on
both sides of the CD. The quantity $\sigma/\rho$ is uniform
everywhere, except across the CD, where it can have a jump.  We denote
the total pressure of intermediate states $p_S^*$, and rest-mass
density left and right of the CD as $\rho_{S,L}$ and $\rho_{S,
  R}$. In the context of internal shocks, if the flow is
ultrarelativistic in the direction of propagation, the velocity
components perpendicular to the flow propagation should be
negligibly small and, hence, they are set up to zero in our
model\footnote{If such velocities were significant, appreciable
  changes in the Riemann structure may result as pointed out in \citet{Aloy:2006p2560} or \citet{Aloy:2008kx}.}.

\subsection{Conditions for the existence of a two-shock solution}
\label{sec:twoshock}

One of the key steps in solving a Riemann problem is to determine
under which conditions internal shocks can form.  States ahead and
behind the shock front are related by the Lichnerowicz adiabat
\citep{Romero:2005zr}
\begin{equation}\label{eq:Lichnerowicz}
(h_b^*)^2 - (h_a^*)^2 - \left(\dsfrac{h_b^*}{\rho_b}-\dsfrac{h_a^*}{\rho_a}\right)(p_b^* - p_a^*) = 0\, .
\end{equation}
Following \citet{Rezzolla:2001ys}, we study the relative velocity
between the states ahead (a) and behind (b) the shock front (all
velocities are measured in the rest frame of the shock, and all 
  thermodynamic properties are measured in the fluid rest frame),
\begin{equation}\label{eq:v_ab}
v_{ab} := \dsfrac{v_a - v_b}{1 - v_a v_b} = 
\sqrt{\dsfrac{(p_b^* - p_a^*)(e_b^* - e_a^*)}{(e_a^* + p_b^*)(e_b^* + p_a^*)}}\, .
\end{equation}
%
%% where $p^*$ is defined in \eref{eq:pstar} 

In our case states ahead of the shock are the initial  (L, R)
states, while states behind the shock are the intermediate
states. Since $v_{ab}$ is Lorentz-invariant, we can measure the
velocity  ahead of the left-propagating ({\it reverse}) shock
  (RS) in the frame in which the CD is at rest,
\begin{equation}\label{eq:v_l}
v_l = \sqrt{\dsfrac{(p_S^* - p_L^*)(e_{S,L}^*(p_S^*) - e_L^*)}{(e_L^* + p_S^*)(e_{S,L}^*(p_S^*) + p_L^*)}}\, .
\end{equation}
 Likewise, the velocity ahead of the right-going ({\it forward})
  shock (FS) measured in the CD frame is 
\begin{equation}\label{eq:v_r}
v_r = -\sqrt{\dsfrac{(p_S^* - p_R^*)(e_{S,R}^*(p_S^*) - e_R^*)}{(e_R^* + p_S^*)(e_{S,R}^*(p_S^*) + p_R^*)}}\, ,
\end{equation}
where $e_{S,L}^*$ and $e_{S,R}^*$ are the energy densities of the
states to the left and to the right of the CD, respectively. The
rest-mass densities $\rho_{S,R}$ and $\rho_{L,R}$ can be obtained from
\eref{eq:Lichnerowicz} and \eref{eq:hstar}.

Since both FS and RS only exist if $p_S^* > p_R^*$ and $p_S^* >
p_L^*$, respectively, with decreasing $p_S^*$ either the FS will
disappear first (for $p_S^* = p_R^* > p_L^*$, giving $v_r=0$) or the
RS will disappear first (for $p_S^* = p_L^* > p_R^*$, giving $v_l =
0$). Using equations \eref{eq:v_l} and \eref{eq:v_r} and the
invariance of the relative velocity between the left and right states,
$v_{lr} := (v_l - v_r)/(1 - v_lv_r)$, we can determine the minimum
relative velocity for which a two-shock solution is possible
\begin{equation}\label{eq:V_LR2S}
(v_{lr})_{2S} = \left\{
  \begin{array}{rl}
    \sqrt{\dsfrac{(p_L^* - p_R^*)(e_{S,R}^*(p_L^*) - e_R^*)}{(e_{S,R}^*(p_L*) + p_R^*)(e_R^* + p_L^*)}} & \mathrm{if}\ p_L^*=p_S^*>p_R^*\\[4mm]
    \sqrt{\dsfrac{(p_R^* - p_L^*)(e_{S,L}^*(p_R^*) - e_L^*)}{(e_{S,L}^*(p_R^*) + p_L^*)(e_L^* + p_R^*)}} & \mathrm{if}\ p_L^*<p_R^*=p_S^*
  \end{array}
  \right.
\end{equation}

Generally, the quantity $(v_{lr})_{2S}$ can be only determined
numerically.  If $(v_{lr}) < (v_{lr})_{2S}$, a single shock and a
rarefaction emerge from the initial discontinuity. It is even possible
that instead of two shocks two rarefactions form (see,
\citealp{Rezzolla:2001ys}).

\section{Energy dissipation efficiency of internal shocks}
\label{sec:edissipation}

Internal shocks in relativistic outflows are invoked as moderately
efficient means of conversion of the kinetic energy of the flow into
radiation. In this section we present our model for inhomogeneous,
ultrarelativistic outflows and provide an operative definition for the
efficiency of conversion of the initial energy of the outflow into
thermal and magnetic energy produced by internal shocks. We assume
that a fraction of this thermal and magnetic energy will be radiated
away.

\subsection{Outflow model}
\label{sec:outflow}

To study internal shocks we idealize interactions of parts of the
outflow moving with different velocities as collisions of homogeneous
shells.  In our model the faster (left) shell catches up with the
slower (right) one yielding, in some cases, a pair of shocks
propagating in opposite directions (as seen from the CD frame).  In
order to cover a wide range of possible flow Lorentz factors and shell
magnetizations, we assume that initially, the flux of energy in the
lab frame is uniform and the same in both shells\footnote{In
  \citet{Mimica:2009qa} we use a similar model to compare afterglow
  ejecta shells with different levels of magnetization, with a slight
  difference that in that study, instead of having the same flux of energy,
  all the ejecta shells were assumed to contain the same total energy.}. The energy
flux for a shell with rest-mass density $\rho$, ratio of thermal
pressure to density $\chi$, magnetization $\sigma$ and Lorentz factor
$\Gamma$ is \citep[e.g.,][]{Leismann:2005rz}.
\begin{equation}\label{eq:F_tau}
F_{\tau}:= \rho \left[ \Gamma^2 (1 + \epsilon + \chi + \sigma) - \Gamma\right] \sqrt{1 - \Gamma^{-2}}\, .
\end{equation}
Using the notation introduced in \secref{sec:initial} and assuming the equality of $F_{\tau}$ in both shells we find that the density contrast $\rho_L$ between left and right shells is
\begin{equation}\label{eq:rho_L}
  \rho_L = \dsfrac{(1+\Delta g)^{-2}\left[1 + \epsilon + \chi + \sigma_R - \Gamma_R^{-1}\right]\sqrt{1 - \Gamma_R^{-2}}}
  {\left[1 + \epsilon + \chi + \sigma_L - \Gamma_R^{-1}(1+\Delta g)^{-1}\right]\sqrt{1 - \Gamma_R^{-2}(1+\Delta g)^{-2}}}\, .
\end{equation}
Considering $\sigma_L$, $\sigma_R$, $\Gamma_R$ and $\Delta g$ as
parameters, we can use \eref{eq:rho_L} to compute the rest of the
variables needed to set-up the Riemann problem. We then compute the
break up of the initial discontinuity between both shells using the
exact Riemann solver developed by \citet{Romero:2005zr}. 

In the following we use a polytropic equation of state with an
adiabatic index $\gad = 4/3$:
\begin{equation}
\epsilon := \dsfrac{p}{(\gad - 1)\rho}
\end{equation}
As we show in the Appendix, the Riemann solver of
  \citet{Romero:2005zr} has been suitably modified to include a more
  realistic equation of state. However, we find no qualitative
  differences between the results using a polytropic EoS (with either
  $\gad=4/3$ or $\gad=5/3$) and $\gad$-variable EoS. Furthermore, the
  quantitative differences are very small in terms of dynamic
  efficiency.

\subsection{Efficiency of energy dissipation by a shock}

To model the dynamic efficiency of energy dissipation we follow
the approach described in \citet{Mimica:2007db}, suitably modified to
account for the fact that in the present work, there can occur
situations where either the FS or the RS do not exist (see
\secref{sec:twoshock}). By using the exact solver we determine the
existence of shocks and (in case one or two shocks exist) obtain the
hydrodynamic state of the shocked fluid. We use subscripts $S, L$ and
$S, R$ to denote shocked portions of left and right shells,
respectively. In the following we treat the efficiency of each shock
separately.

\subsubsection{Reverse shock}

To compute the efficiency we need to compare the energy content of the
initial (unshocked) faster shell with that of the shocked shell at the
moment when RS has crossed the initial shell. Assuming an initial
shell width $\Delta x$, we define total initial kinetic, thermal and
magnetic energy (see also equations (A.1) - (A.3) of
\citealt{Mimica:2007db})
\begin{equation}\label{eq:energies}
  \begin{array}{rcl}
    E_K(\Gamma, \rho, \Delta x) & := & \Gamma (\Gamma - 1) \rho \Delta x\\[4mm]
    E_T (\Gamma, \rho, p, \Delta x)& := & [(\rho \varepsilon + p) \Gamma^2 - p] \Delta x\\[4mm]
    E_M (\Gamma, \rho, \sigma, \Delta x) & := & \left(\Gamma^2 - \dsfrac{1}{2}\right)\rho \sigma \Delta x
  \end{array}\, .
\end{equation}
When the RS crosses the whole initial shell, the length of the
compressed shell (i.e., the fluid between the RS and the CD) is
$\zeta_L \Delta x$, where
\[
\zeta_L: =\dsfrac{v_{CD} - v_{S, L}}{v_L - v_{S, L}} < 1
\]
and $v_{CD}$ and $v_{S, L}$ are velocities (in the lab frame) of the
contact discontinuity and the RS, both obtained from the solver.
Without loss of generality, we can normalize the initial shell width
so that $\Delta x=1$. Then we define the dynamic \emph{thermal}
efficiency
\begin{equation}\label{eq:thermalL}
  \varepsilon_{T, L} := \dsfrac{E_T(\Gamma_{S, L}, \rho_{S, L}, p_{S, L}, \zeta_L) -
    E_T(\Gamma_R (1 + \Delta g), \rho_L, \chi \rho_L, 1)}{E_0}\, ,
\end{equation}
and the dynamic \emph{magnetic} efficiency
\begin{equation}\label{eq:magneticL}
  \varepsilon_{M, L}:= \dsfrac{E_M(\Gamma_{S, L}, \rho_{S, L}, \sigma_{S, L}, \zeta_L) - E_M(\Gamma_R (1 + \Delta g), \rho_L, \sigma_L, 1)}{E_0}\, ,
\end{equation}
where $E_0$ is the total initial energy of both shells 
\begin{equation}\label{eq:E_0}
  \begin{array}{rl}
    E_0 &:= E_K(\Gamma_R  (1 + \Delta g), \rho_L, 1) + E_T(\Gamma_R (1 + \Delta g), \rho_L, \chi \rho_L, 1)\\[4mm]
    &+ E_M(\Gamma_R (1 + \Delta g), \rho_L, \sigma_L, 1) + E_K(\Gamma_R, 1, 1) \\[4mm]
    &+ E_T(\Gamma_R, 1, \chi, 1) + E_M(\Gamma_R, 1, \sigma_R, 1)\, .
    \end{array}\, 
\end{equation}

Equations \eref{eq:thermalL} and \eref{eq:magneticL} express the
fraction of the initial energy that the RS has converted into thermal
and magnetic energy, respectively.

\subsubsection{Forward shock}

In complete analogy, we define the thermal and magnetic
efficiencies for the forward shock,
\begin{equation}\label{eq:thermalR}
  \varepsilon_{T, R} := \dsfrac{E_T(\Gamma_{S, R}, \rho_{S, R}, p_{S, R}, \zeta_R) - E_T(\Gamma_R, 1, \chi, 1)}{E_0}\, ,
\end{equation}
\begin{equation}\label{eq:magneticR}
  \varepsilon_{M, R}:= \dsfrac{E_M(\Gamma_{S, R}, \rho_{S, R}, \sigma_{S,
      R}, \zeta_R) - E_M(\Gamma_R, 1, \sigma_R, 1)}{E_0}\, ,
\end{equation}
where
\[
\zeta_R := \dsfrac{v_{S, R} - v_{CD}}{v_{S, R} - v_R} < 1\, .
\]
$v_{S, R}$ is the velocity of the FS in the lab frame. Here we set
$\varepsilon_{T, R}=\varepsilon_{M, R}=0$ if the forward shock is absent.

%\subsubsection{Dynamic energy dissipation efficiency}

Combining equations \eref{eq:thermalL}, \eref{eq:magneticL},
\eref{eq:thermalR} and \eref{eq:magneticR} we define the
dynamic thermal and magnetic efficiency of internal shocks
\begin{equation}\label{eq:thermal}
  \varepsilon_T := \varepsilon_{T, L} + \varepsilon_{T, R}
\end{equation}
\begin{equation}\label{eq:magnetic}
  \varepsilon_M:= \varepsilon_{M, L} + \varepsilon_{M, R}\, .
\end{equation}
%
%We define the total initial energy in both shells as

We point out that these definitions of efficiency generalize the ones
typically used when cold, unmagnetized shell collisions are
considered. In that case, initially one only has bulk kinetic energy
in the shells (i.e., $E_0=E_K(\Gamma_L, \rho_L, 1) + E_K(\Gamma_R, 1,
1)$). In case of collisions of arbitrarily magnetized shells with
arbitrarily initial thermal content, $E_0$ can be substantially larger
than the initial kinetic energy in the shells.

\subsection{Efficiency of energy dissipation by a rarefaction}

In a rarefaction there is a net conversion of magnetic and/or thermal
energy into kinetic energy, thus the net dynamic efficiency produced
by a rarefaction, defined as in e.g., Eq.~\eref{eq:magneticL}, should
be negative. The consequence of this is that, when we obtain a
shock-contact-rarefaction or rarefaction-contact-shock structure as a
solution to the Riemann problem, it may happen that the total (left
plus right) thermal or magnetic efficiency
(Eqs.~\eref{eq:thermal}-\eref{eq:magnetic}) was negative. However,
this situation does not correctly model the fact that, in cases where
a shock exists only in one of the shells, it is still able to radiate
away part of the thermal or magnetic energy behind it, even though
there is a rarefaction propagating through the other shell. We
  also point out that rarefactions happen in our case, where we model
  initially cold flows, as a result of a net conversion of magnetic
  into kinetic energy. This kinetic energy can be further recycled by
  the flow, and dissipated in the course of ulterior binary
  collisions.  Therefore, directly summing the (positive) dynamic
efficiency of conversion of kinetic-to-thermal/magnetic energy in a
shock to the (negative) dynamic efficiency of conversion in a
rarefaction is inadequate. The total dynamic efficiency in a case
where only one shock forms will be determined only by the efficiency
in the shocked shell. Thus, we set $\varepsilon_{T, L} =
\varepsilon_{M, L}=0$ ($\varepsilon_{T, R} = \varepsilon_{M, R}=0$) if
the reverse (forward) shock is absent. We note that this contrasts
with previous works on internal collisions of unmagnetized shells
\citep[e.g.,][]{Kino:2004p811}, and may yield higher values of the net
dynamic efficiency.

\section{Parametric study of the dynamic dissipation efficiency}
\label{sec:parametric}

Next we study the dynamic dissipation efficiency in the process of
collision of cold, magnetized shells.The shells are
  assumed to be cold because in the standard fireball model \citep[e.g.,][]{Piran:2005p2632}
, almost all the internal energy of the ejecta has been
  converted to kinetic energy {\it before} internal shocks start to
  show up. Thus, a regime where the parameter $\chi$ is large does not
  properly model an efficiently accelerated ejecta by non-magnetic
  processes. On the other hand, if the ejecta were accelerated by
  magnetic fields (like in Poynting-dominated flow models; e.g., \citealt{Usov:1992hp}), then the flow is cold all the way from the beginning to the
  internal shock phase, and then $\chi$ should also be small in such a
  case.

  For all the models in this paper, in order to reduce the dimensions
  of the parameter space, we fix $\chi = 10^{-4}$ uniform everywhere,
to model initially cold shells and, unless otherwise specified, set
$\Delta g = 1$ as a reference value. We choose $\chi$
  sufficiently small so that it does not influence the solution of the
  Riemann problem.  In the first two subsections we consider blazar
and GRB regimes.  Then, we study the flow structure for three
representative Riemann problems, and end the section with a discussion
of the impact of varying $\Delta g$ on the efficiency.

\subsection{Blazar regime}

\begin{figure}
\includegraphics[width=8.5cm]{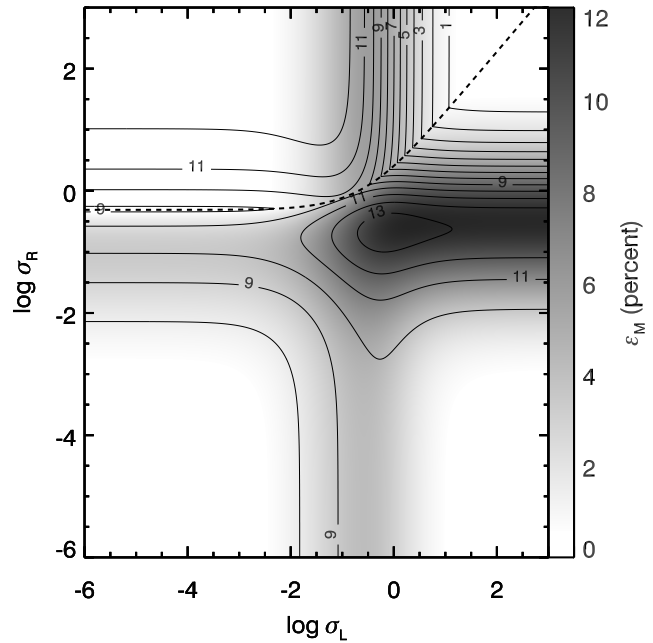}
\caption{Contours: total dynamic efficiency $\varepsilon_T +
  \varepsilon_M$ (eqs.~\eref{eq:thermal} and \eref{eq:magnetic}) in the
  blazar regime ($\Gamma_R = 10$, $\Delta g = 1$) for different
  combinations of $(\sigma_L, \sigma_R)$. Contours indicate the
  efficiency in percent and their levels are $1$, $2$, $3$, $4$, $5$,
  $6$, $7$, $8$, $9$, $10$, $11$, $12$ and $13$. In the region of the
  parameter space above the dashed line there is no forward shock,
  while the reverse shock is always present for the considered
  parametrization. Filled contours: magnetic efficiency $\varepsilon_M$
  in percent.}
\label{fig:blazar}
\end{figure}

In the blazar regime we set $\Gamma_R = 10$ as a typical value of the
Lorentz factor of the outflowing material. We continuously vary
$\sigma_L$ and $\sigma_R$ and show contours of total efficiency
($\varepsilon_T$ + $\varepsilon_M$) in \figref{fig:blazar}.

%%%
%%%

The maximum efficiency is attained for moderately magnetized slower
shells ($\sigma_R\approx 0.2$) and highly magnetized left shells
($\sigma_L \approx 1$). The broad region to the right of the
efficiency maximum is independent of $\sigma_L$ because in a collision
with such a highly magnetized fast shell almost all the energy is
dissipated by the FS. In the region above the dashed line of
\figref{fig:blazar} the FS is absent and, thus, since only the RS
dissipates the initial energy, the efficiency slightly drops. However,
the transition between the regime where the two shocks operate or only
the RS exists is smooth. The reason being that the efficiency below
the separatrix of the two regimes and close to it is dominated by the
contribution of RS.

As expected, when either $\sigma_L$ or $\sigma_R$ approach low values,
the dynamic efficiency ceases to depend on them. This can be seen in
the center of the left side of \figref{fig:blazar} where, for
$\sigma_L\ll 1$, the dynamic efficiency only depends on
$\sigma_R$. The converse is true in the center of the lower side of
the figure, where $\sigma_R\ll 1$.

\subsection{GRB regime}

%%%
\begin{figure}
\includegraphics[width=8.5cm]{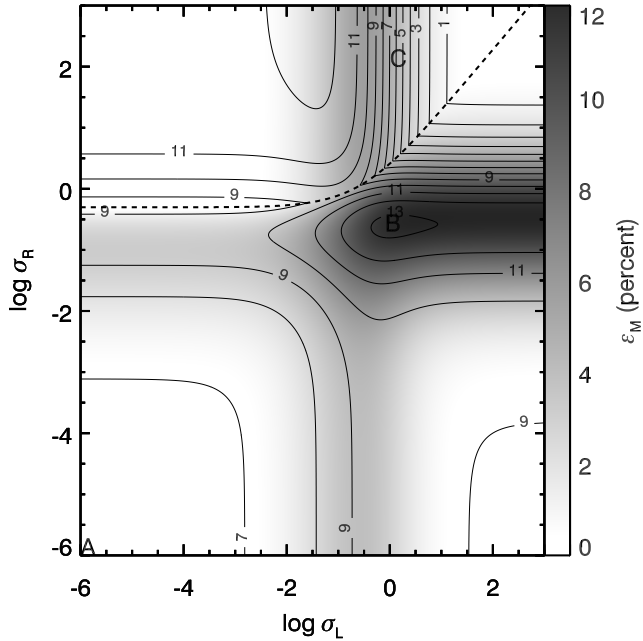}
\caption{Same as \figref{fig:blazar}, but in the GRB regime ($\Gamma_R = 100$, $\Delta g = 1$).}
\label{fig:grb}
\end{figure}
%%%

The results in the GRB regime ($\Gamma_R = 100$, $\Delta g = 1$) are
shown in \figref{fig:grb}. The general shape of the contours is
similar to \figref{fig:blazar}, which is expected since both in the
blazar and in the GRB regime the flow is utrarelativistic and most of
the quantities which depend on the \emph{relative} velocity between
the faster and the slower shell depend only weakly on $\Gamma_R$,
$\Delta g$ being the crucial parameter \citep{Daigne:1998wq}. For
example, the dashed curve which delimits regions with and without a
forward shock does not depend on $\Gamma_R$ but only on $\Delta g$.

The maximum of the dynamic efficiency in the GRB regime is localized
at roughly the same spot as in the blazar regime. However, compared to
the former case the region of maximum efficiency (i.e., where
$\varepsilon_T+\varepsilon_M \gtrsim 0.13$) is smaller.

\subsection{Flow structure}
\label{sec:flowstr}

\begin{figure}
\includegraphics[width=8.5cm]{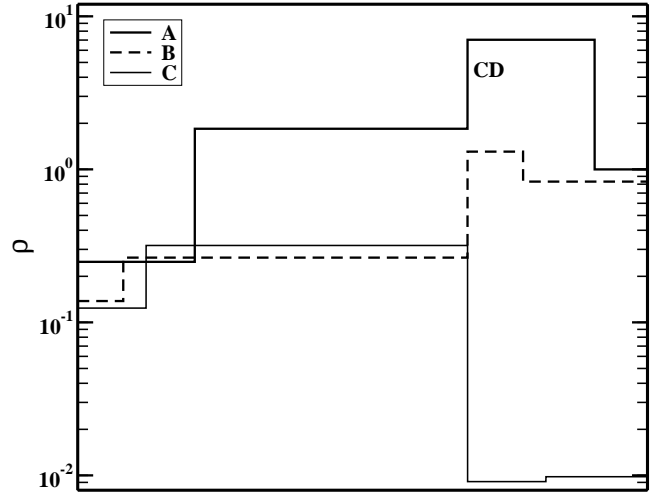}
\caption{Rest-mass density profile of models $A$, $B$ and $C$ (see
  legends) whose parameters are given in
  Sect.~\ref{sec:flowstr}. Profiles have been shifted so that the CD
  for all models coincides. In models $A$ and $B$ the FS and the RS
  are clearly visible, while in the model $C$ a rarefaction wave is
  visible as a small ``step'' to the right of the CD.}
\label{fig:structure}
\end{figure}

In this subsection we study the flow structure for three
representative models in the GRB regime. Their location in the
parameter space is marked by letters $A$, $B$ and $C$ in
\figref{fig:grb}. Model $A$ corresponds to a prototype of interaction
between non-magnetized shells $(\sigma_L = \sigma_R = 10^{-6})$. Model
$B$ is picked up to illustrate the flow structure at the maximum
efficiency $(\sigma_L = 0.8,\ \sigma_R = 0.2)$. Finally, model $C$
corresponds to the case when the FS is absent $(\sigma_L = 1,\
\sigma_R = 10^2)$. We show the rest mass density profile of these
models in \figref{fig:structure}.
%%%
%%%
Model $A$ (thick full line on \figref{fig:structure}) shows two strong
shocks which dissipate kinetic into thermal energy. In contrast, model
$B$ (dashed line) has much weaker shocks due to non-negligible
magnetization in both shells. Finally, model $C$ (thin full line) does
not have forward shock due to very high magnetization in the slower
shell.

All three models have a substantial dynamic efficiency, but there is a
qualitative difference among them. In model $A$ internal shocks
dissipate kinetic to thermal energy only (thermal efficiency). In
model $B$ shocks mainly compress the magnetic field (magnetic
efficiency) and dissipate only a minor fraction of the initial kinetic
and magnetic energy to thermal energy. Finally, in the model $C$ only
the reverse shock is active, compressing the faster shell magnetic
field.

\subsection{Dependence on $\Delta g$ and on $\Delta s$}
\label{sec:deltag-deltas}

%%%
\begin{figure}
\includegraphics[width=8.5cm]{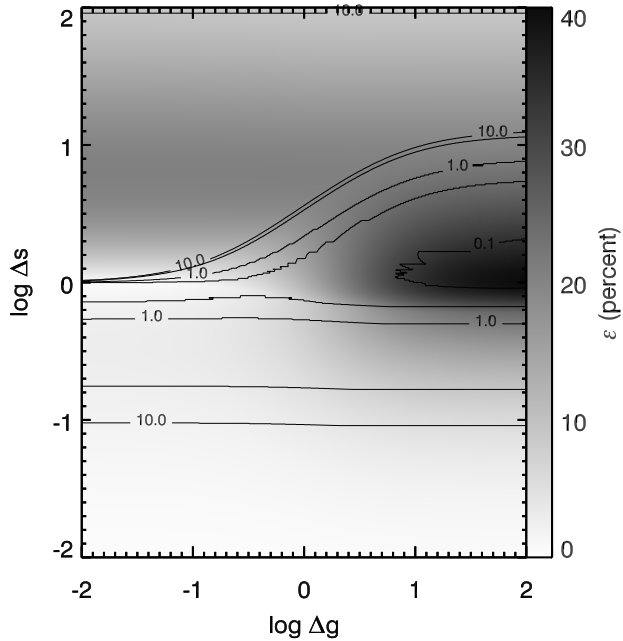}
\caption{The gray scale indicates the value of the maximum total
  dynamic efficiency (in percent) as a function of the parameter
  pair $(\Delta g, \Delta s)$. The values of the rest of the
  parameters are fixed to $\Gamma_R=100$, and $\chi=10^{-4}$.
  Contours: magnetization of the slowest shell: $\sigma_R=0.1$, $0.5$,
  1, 5 and 10.}
\label{fig:deltag-deltas}
\end{figure}
%%%
The choice of a relatively small value of $\Delta g$ in the previous
section is motivated by the results of numerical simulations of
relativistic outflows
\citep[e.g][]{Aloy:2000ad,Aloy:2005zp,Mizuta:2006p1126,Zhang:2003p1193,Zhang:2004p1195,
  Morsony:2007p1208,Lazzati:2009p1210,Mizuta:2009p1213} where $\Delta
g < 2$ between adjacent parts of the flow that may catch up (but see,
e.g., \citealt{Kino:2008p1217}, who find $\Delta g\sim 1-19$
appropriate to model Mrk 421). This adjacent flow regions can be
assimilated to pairs of shells whose binary collision we are
considering here.

However, it has been confirmed by several independent works (KPS97;
\citealt{Beloborodov:2000p632}; \citealt{Kobayashi:2000p599};
\citealt{Kino:2004p811}, etc.)  that, in order to achieve a high
efficiency (more than a few percent) in internal collisions of
unmagnetized shells, the ratio between the maximum ($\Gamma_{\rm
  max}$) and the minimum ($\Gamma_{\rm min}$) Lorentz factor of the
distribution of initial shells should be $\Gamma_{\rm max}/\Gamma_{\rm
  min}>10$.

In view of these results, we have also made an extensive analysis of
the dependence of the dynamic efficiency on the variation of $\Delta
g$. Since we are also interested in evaluating the influence of the
magnetic fields on the results, we define a new variable
\begin{equation}\label{eq:magnetics}
  \Delta s= \frac{1+\sigma_L}{1+\sigma_R}\, ,
\end{equation}
and plot (Fig.~\ref{fig:deltag-deltas}) the value of the maximum
efficiency reached for every combination $(\Delta g, \Delta s)$ and
fixed values of the rest of the parameters to $\Gamma_R=100$, and
$\chi=10^{-4}$. To be more precise, for fixed $\Delta g$ and $\Delta
s$ we need to look for the maximum of the efficiency of all models
whose $\sigma_L$ and $\sigma_R$ satisfy Eq.~\eref{eq:magnetics}.  It
is evident from Fig.~\ref{fig:deltag-deltas} that the maximum total
dynamic efficiency grows (non-monotonically) with increasing $\Delta
g$, in agreement with the above mentioned works (where unmagnetized
collisions have been considered). Indeed, a large value $\Delta g
\gtrsim 10$ yields dynamic efficiency values $\sim 40\%$ if both
shells are moderately magnetized ($\sigma_R\sim\sigma_L \lesssim
0.1$). Nevertheless, the amount of increase of efficiency with $\Delta
g$ depends strongly on $\Delta s$. For $|\Delta s| \gtrsim 1$,
corresponding to cases where the slower shell is highly magnetized
($\sigma_R\gtrsim 4$), the maximum dynamic efficiency is almost
independent of $\Delta g$; while for $|\Delta s| \lesssim 1$, the
maximum dynamic efficiency displays a strong, non-monotonic dependence
on $\Delta g$.

It is remarkable that values $5 \lesssim \Delta s \lesssim 100$ yield
dynamic efficiencies in excess of $\sim 20\%$, regardless of the
relative Lorentz factor between the two shells. In this region of the
parameter space the maximal dynamic efficiency happens when both
shells are magnetized ($\sigma_R>10$, $\sigma_L>50$), and the total
dynamic magnetic efficiency dominates the total dynamic efficiency.

\section{Discussion}
\label{sec:discussion}

We have focused in this paper on the estimation of the dynamic
efficiency of conversion of kinetic-to-thermal/magnetic energy in
collisions (internal shocks) of magnetized shells in relativistic
outflows.  A fundamental difference between the internal collisions in
magnetized and unmagnetized outflows is the fact that in the former
case not only shocks but also rarefactions can form. Thus, one would
naturally expect a reduced dynamic efficiency in the magnetized
case. However, we have shown that such dynamic efficiency may reach
values $\sim 10\%-40\%$, in a wide range of the parameter space
typical for relativistic outflows of astrophysical interest (blazars
and GRBs). Thus, the dynamic efficiency of moderately magnetized shell
interactions is larger than in the corresponding unmagnetized
case. This is because when the shells are moderately magnetized, most
of the initial shell kinetic energy is converted to magnetic energy,
rather than to thermal energy.

The difference in efficiency between flows with moderate Lorentz
factors ($\Gamma_R=10$) and ultrarelativistic ones ($\Gamma_R=100$) is
very small, because in the ultrarelativistic kinematic limit (i.e.,
$\Gamma \gg 1$), once the energy flux and the magnetizations of both
shells are fixed, the key parameter governing the dynamic
efficiency is $\Delta g$ rather than $\Gamma_R$. From numerical
simulations one expects that any efficiently accelerated outflow will
not display huge variations in the velocity between adjacent regions
of flow. Therefore, values of $\Delta g\simeq 1$ seem reasonable and
$\Delta g =1$ has been taken as a typical value for both blazars and
GRB jets. A fixed value of $\Delta g=1$ brings maximum efficiency when
the magnetizations of the colliding shells are
$(\sigma_L,\sigma_R)\simeq(1,0.2)$. Larger dynamic efficiency values
$\sim 40\%$ are reached by magnetized internal shocks if $\Delta
g\gtrsim 10$ and $|\Delta s|\lesssim 1$, corresponding to cases
where the magnetization of both shells is moderate ($\sigma_R\simeq
\sigma_L\lesssim 0.1$).

Consistent with our previous work \citep{Mimica:2007db}, in the limit
of low magnetization of both shells, the kinetic energy is mostly
converted into thermal energy, where the increased magnetic energy in
the shocked plasma is only a minor contribution to the total dynamic
efficiency, i.e., $\varepsilon_T \ll \varepsilon_M$. Here we find that
as the magnetization of the shells grows, the roles of $\varepsilon_T$
and $\varepsilon_M$ are exchanged, so that $\varepsilon_T <
\varepsilon_M$ (at the maximum dynamic efficiency $\varepsilon_T
\simeq 0.1 \varepsilon_M$). If the magnetization of both shells is
large, the dynamic efficiency decreases again because producing shocks
in highly magnetized media is very difficult. All these
  conclusions are independent on the EoS used to model the plasma,
  i.e., they are both qualitatively and quantitatively basically the
  same independent on whether a polytropic EoS with a fixed adiabatic
  index is taken (either $\gad=4/3$ or $\gad=5/3$) or a more general,
  analytic approximation to the exact relativistic EoS (the \emph{TM}
  EoS; see Appendix) is considered.

The comparison of our results with previous analytic or semi-analytic
works (e.g.,KPS97; \citealt{Beloborodov:2000p632};
\citealt{Kobayashi:2001p618}; \citealt{Kobayashi:2002p1216}) is not
straightforward. Generally, these works compute the efficiency of the
collision of shells without computing their (magneto-)hydrodynamic
evolution and, on the other hand, these works include not only a
single collision, but the multiple interactions of a number of dense
shells. The bottom line in these previous works is that internal
collisions of unmagnetized shells can be extremely efficient; the
efficiency exceeding $40\%$, or even $\sim 100\%$
\citep{Beloborodov:2000p632} if the spread of the Lorentz factor
(i.e., the ratio between the Lorentz factor of the faster,
$\Gamma_{\rm max}$, and of the slower $\Gamma_{\rm min}$ shell in the
sample) is large ($\Gamma_{\rm max} /\Gamma_{\rm min}=10^3$; e.g.,
KPS97, \citealt{Kobayashi:2002p1216}).  For a more moderate spread of
the Lorentz factor $\Gamma_{\rm max}/\Gamma_{\rm min} =10$, the
efficiency is $\sim 20\%$. We note that these high efficiencies are
reached because a large number of binary collisions is included in the
model (not only a single one as in our case). Thus, the kinetic energy
which is not dissipated in the first generation of collisions (between
the initially set up shells), can be further converted into internal
energy as subsequent generations of collisions take place. In
contrast, we find that moderate magnetizations of both shells
($\sigma\lesssim 0.1$) and $\Delta g\gtrsim 10$ (which would roughly
correspond to $\Gamma_{\rm max}/\Gamma_{\rm min} =9$) are enough for a
single binary collision to reach a total dynamic efficiency of $\sim
40\%$.

We point out that the energy radiated in the collision of magnetized
shells is only a fraction, $f_r\simeq 0.1$
\citep[e.g.,][]{Panaitescu:1999p1022,Kumar:1999p1084} to $f_r\simeq 1$
\citep[e.g.,][]{Beloborodov:2000p632} of the energy dynamically
converted into thermal or magnetic energy. Thus, the radiative
efficiency of the process, measured as the fraction of the total
initial energy converted into radiation, will be $1/f_r$ times smaller
than the computed dynamic efficiency. Even considering this factor,
single binary collisions between moderately magnetized shells may
yield efficiencies $\sim 0.4f_r$, which can obviously rise if many
binary collisions take place in the flow reprocessing the remaining
kinetic energy of the first generation of interacting shells (in the
same statistical way as discussed by
\citealt{Kobayashi:2000p599}). Therefore, on the light of our
  results, binary collisions in relativistic magnetized flows are
  efficient enough, from the dynamical point of view, to be a valid
  mechanism to dissipate the bulk kinetic energy of relativistic
  ejecta. Hence, the main restriction on the radiative efficiency
  comes from the radiation mechanism setting the limiting factor
  $f_r$.

We stress that we are not assuming any particular radiation mechanism
in this study (thus, determining a value for $f_r$). Therefore,
we compute the dynamic efficiency including not only the increased
thermal energy in the flow, but also the extra magnetic energy
resulting from the magnetic field compression. This is justified
because, although standard shock acceleration mechanisms are
inefficient in very magnetized shocks
\citep[e.g.,][]{Sironi:2009p414}, other mechanisms might extract the
energy from the whole volume of a very magnetized fluid
\citep[e.g.,][]{Thompson:1994p492,Giannios:2007by}.

The estimated dynamic efficiency in the binary collision of
magnetized shells will be completed in a future work by accounting for
the numerical MHD evolution of such building blocks of the internal
shock models. A step further would be to compute the radiative
efficiency using the method devised in \cite{Mimica:2009p1445}.

\section*{Acknowledgments}
MAA is a Ram\'on y Cajal Fellow of the Spanish Ministry of Education
and Science. We acknowledge the support from the Spanish Ministry of
Education and Science through grants AYA2007-67626-C03-01 and
CSD2007-00050. We thank Jos\'e-Mar\'{\i}a Mart\'{\i} and
Jos\'e-Mar\'{\i}a Ib\'a\~{n}ez for their support and critical
discussions. The authors thankfully acknowledge the computer
resources, technical expertise and assistance provided by the
Barcelona Supercomputing Center - Centro Nacional de
Supercomputaci\'on.

\appendix

\section{Equation of state}
\label{sec:EoS}

In this Appendix we discuss the effects of using a more realistic
equation of state on our results. We choose the \emph{TM} analytic
approximation to the Synge equation of state (EoS) 
\citep{deBerredoPeixoto:2005p2625,Mignone:2005p2379}. In the \emph{TM} EoS the specific enthalpy
can be written as (using the notation of \secref{sec:initial})
\begin{equation}\label{eq:htm}
  h^*_{TM} := \dsfrac{5}{2}\dsfrac{p^*}{\rho} - \dsfrac{\sigma}{4} +
  \left[\dsfrac{9}{4}\left(\dsfrac{p^*}{\rho} - \dsfrac{\sigma}{2}\right)^2 + 1\right]^{1/2}\, .
\end{equation}
and the specific internal energy
\begin{equation}\label{eq:epstm}
  \epsilon_{TM}:= \dsfrac{3}{2}\dsfrac{p}{\rho} + \left[\dsfrac{9}{4}\left(\dsfrac{p}{\rho}\right)^2 + 1\right]^{1/2} - 1\, .
\end{equation}
From \eref{eq:htm} it can be seen that the limit $\sigma = 0$ the
effective adiabatic index of this EoS lies between $4/3$ and $5/3$. We
modified the \citet{Romero:2005zr} solver to include the \emph{TM}
EoS.

%%%
\begin{figure}
\includegraphics[width=8.5cm]{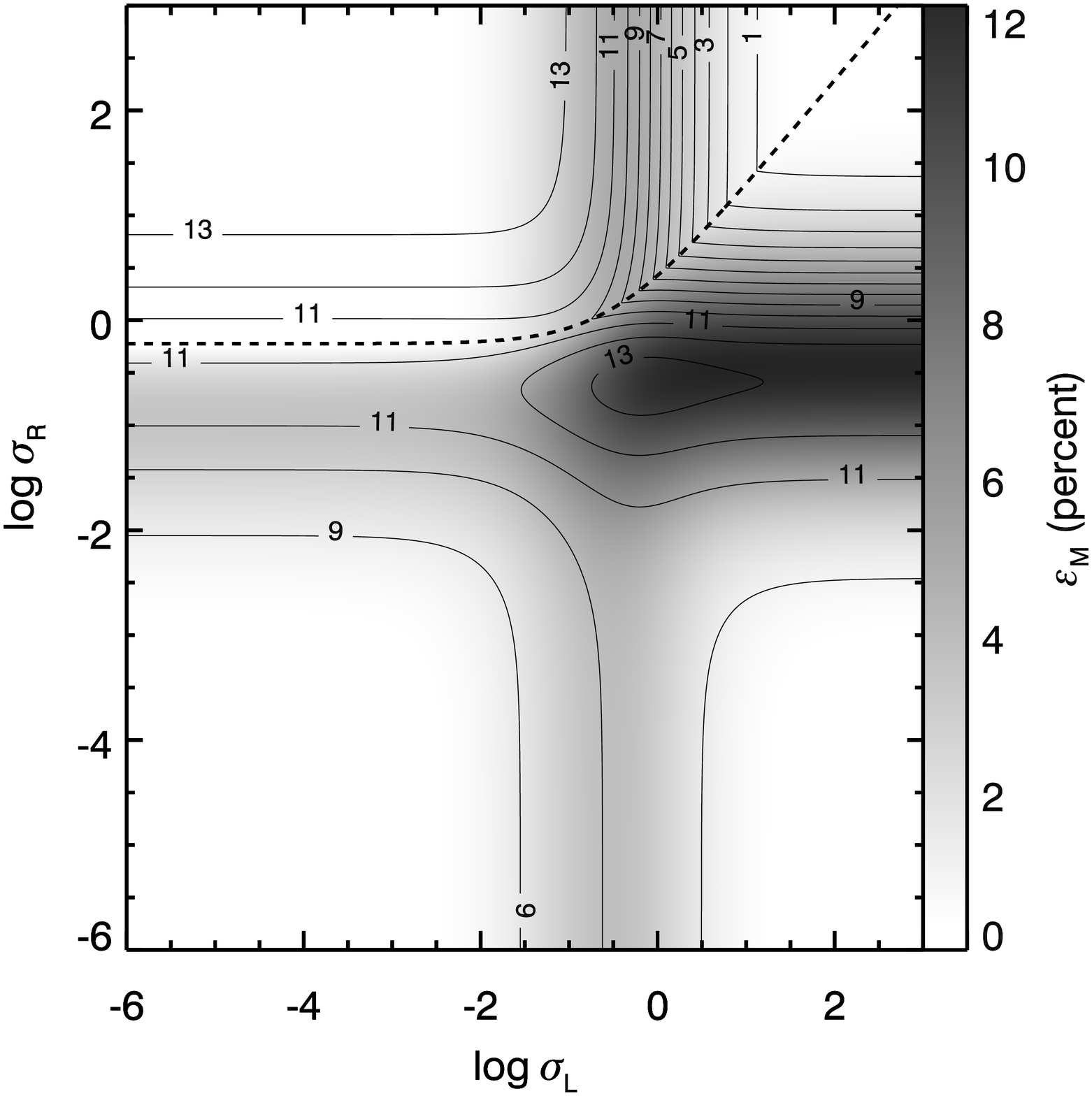}
\caption{Same as \figref{fig:grb}, but for the \emph{TM} equation of state.}
\label{fig:TMgrb}
\end{figure}
%%%

\begin{figure}
\includegraphics[width=8.5cm]{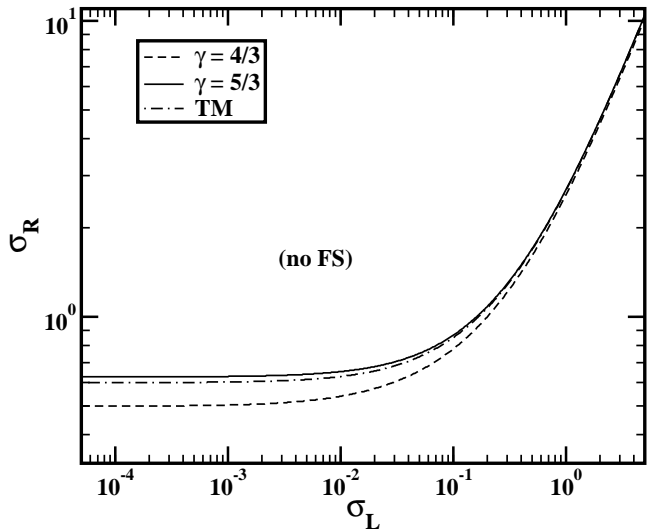}
\caption{Comparison of the separatrix lines dividing regions of
  existence and non-existence of the forward shock (located above each
  line, the region where the forward shock is not formed). Solid and
  dashed lines correspond to the polytropic EoS with $\gad = 5/3$ and
  $\gad = 4/3$, respectively, while the dot-dashed line shows result
  for the \emph{TM} EoS. }
\label{fig:FS}
\end{figure}

On \figref{fig:TMgrb} we show the dynamical efficiency in the GRB
regime using the \emph{TM} EoS. Comparison of \figref{fig:grb} and
\figref{fig:TMgrb} shows that, overall, the dynamical efficiency is
higher when using the \emph{TM} EoS, but the qualitative features
remain the same in both cases. Also, as expected, in the highly
magnetized regime the differences are minor, since in both cases
(polytropic or \emph{TM} EoS) the effective adiabatic index approaches
2 in such a regime. Figure~\ref{fig:FS} shows the influence of the EoS
on the existence of the FS. The only difference between models with
different EoS appears in the region where the faster shell is weakly
magnetized. There a slightly higher (lower) magnetization of the
slower shell is needed to suppress the FS when using \emph{TM} than
when using a polytropic EoS with $\gad=4/3$ ($\gad = 5/3$). The
overlap of all three curves in \figref{fig:FS} in the limit of
high magnetization of both shells, shows again the result that the
choice of EoS plays no role in such a regime, as expected. We note
that the separatrix between the regions of existence and non-existence
of the FS corresponding to the case $\gad=5/3$ lies closer to that of
the \emph{TM} EoS than that corresponding to the case $\gad=4/3$. This
is a natural consequence of the fact that the initial shells are both
cold, and thus, the effective adiabatic index is closer to $5/3$ than
to $4/3$, at low magnetizations.

%%%
%%%

We note that in the case of the external shocks, where relativistic
fluid encounters a cold external medium the choice of the realistic
EoS can influence results much more dramatically than in the case of
the internal shocks \citep[see e.g.,][]{Meliani:2008p2303}.

\bibliographystyle{mn2e}
\bibliography{magint.bib}

\begin{thebibliography}{}

\bibitem[\protect\citeauthoryear{Aloy, Janka \& M{\"u}ller}{Aloy
  et~al.}{2005}]{Aloy:2005zp}
Aloy M.~A.,  Janka H.-T.,    M{\"u}ller E.,  2005, \aap, 436, 273

\bibitem[\protect\citeauthoryear{Aloy \& Mimica}{Aloy \&
  Mimica}{2008}]{Aloy:2008kx}
Aloy M.~A.,  Mimica P.,  2008, \apj, 681, 84

\bibitem[\protect\citeauthoryear{Aloy, M{\"u}ller, Ib{\'a}{\~n}ez, Mart\'{\i}
  \& MacFadyen}{Aloy et~al.}{2000}]{Aloy:2000ad}
Aloy M.~A.,  M{\"u}ller E.,  Ib{\'a}{\~n}ez J.~M.,  Mart\'{\i} J.,    MacFadyen
  A.,  2000, \apjl, 531, L119

\bibitem[\protect\citeauthoryear{Aloy \& Rezzolla}{Aloy \&
  Rezzolla}{2006}]{Aloy:2006p2560}
Aloy M.~A.,  Rezzolla L.,  2006, \apj, 640, L115

\bibitem[\protect\citeauthoryear{Beloborodov}{Beloborodov}{2000}]{Beloborodov:%
2000p632}
Beloborodov A.~M.,  2000, \apj, 539, L25

\bibitem[\protect\citeauthoryear{Bo{\v s}njak, Daigne \& Dubus}{Bo{\v s}njak
  et~al.}{2009}]{Bosnjak:2009p1427}
Bo{\v s}njak {\v Z}.,  Daigne F.,    Dubus G.,  2009, \aap, 498, 677

\bibitem[\protect\citeauthoryear{Daigne \& Mochkovitch}{Daigne \&
  Mochkovitch}{1998}]{Daigne:1998wq}
Daigne F.,  Mochkovitch R.,  1998, \mnras, 296, 275

\bibitem[\protect\citeauthoryear{de Berredo-Peixoto, Shapiro \&
  Sobreira}{de~Berredo-Peixoto et~al.}{2005}]{deBerredoPeixoto:2005p2625}
de Berredo-Peixoto G.,  Shapiro I.~L.,    Sobreira F.,  2005, Modern Physics
  Letters A, 20, 2723

\bibitem[\protect\citeauthoryear{Fan \& Piran}{Fan \&
  Piran}{2006}]{Fan:2006p1375}
Fan Y.,  Piran T.,  2006, \mnras, 369, 197

\bibitem[\protect\citeauthoryear{Fan, Wei \& Zhang}{Fan
  et~al.}{2004}]{Fan:2004p1007}
Fan Y.~Z.,  Wei D.~M.,    Zhang B.,  2004, \mnras, 354, 1031

\bibitem[\protect\citeauthoryear{Giacomazzo \& Rezzolla}{Giacomazzo \&
  Rezzolla}{2006}]{Giacomazzo:2006p2513}
Giacomazzo B.,  Rezzolla L.,  2006, Journal of Fluid Mechanics, 562, 223

\bibitem[\protect\citeauthoryear{Giannios \& Spruit}{Giannios \&
  Spruit}{2007}]{Giannios:2007by}
Giannios D.,  Spruit H.~C.,  2007, \aap, 469, 1

\bibitem[\protect\citeauthoryear{Kennel \& Coroniti}{Kennel \&
  Coroniti}{1984}]{Kennel:1984kx}
Kennel C.~F.,  Coroniti F.~V.,  1984, \apj, 283, 694

\bibitem[\protect\citeauthoryear{Kino, Mizuta \& Yamada}{Kino
  et~al.}{2004}]{Kino:2004p811}
Kino M.,  Mizuta A.,    Yamada S.,  2004, \apj, 611, 1021

\bibitem[\protect\citeauthoryear{Kino \& Takahara}{Kino \&
  Takahara}{2008}]{Kino:2008p1217}
Kino M.,  Takahara F.,  2008, \mnras, 383, 713

\bibitem[\protect\citeauthoryear{Kobayashi, Piran \& Sari}{Kobayashi
  et~al.}{1997}]{Kobayashi:1997p657}
Kobayashi S.,  Piran T.,    Sari R.,  1997, \apj, 490, 92

\bibitem[\protect\citeauthoryear{Kobayashi, Ryde \& MacFadyen}{Kobayashi
  et~al.}{2002}]{Kobayashi:2002p1216}
Kobayashi S.,  Ryde F.,    MacFadyen A.,  2002, \apj, 577, 302

\bibitem[\protect\citeauthoryear{Kobayashi \& Sari}{Kobayashi \&
  Sari}{2000}]{Kobayashi:2000p599}
Kobayashi S.,  Sari R.,  2000, \apj, 542, 819

\bibitem[\protect\citeauthoryear{Kobayashi \& Sari}{Kobayashi \&
  Sari}{2001}]{Kobayashi:2001p618}
Kobayashi S.,  Sari R.,  2001, \apj, 551, 934

\bibitem[\protect\citeauthoryear{Kumar}{Kumar}{1999}]{Kumar:1999p1084}
Kumar P.,  1999, \apj, 523, L113

\bibitem[\protect\citeauthoryear{Lazzati, Ghisellini \& Celotti}{Lazzati
  et~al.}{1999}]{Lazzati:1999p1360}
Lazzati D.,  Ghisellini G.,    Celotti A.,  1999, \mnras, 309, L13

\bibitem[\protect\citeauthoryear{Lazzati, Morsony \& Begelman}{Lazzati
  et~al.}{2009}]{Lazzati:2009p1210}
Lazzati D.,  Morsony B.~J.,    Begelman M.~C.,  2009, \apjl, 700, L47

\bibitem[\protect\citeauthoryear{Leismann, Ant{\'o}n, Aloy, M{\"u}ller,
  Mart\'{\i}, Miralles \& Ib{\'a}{\~n}ez}{Leismann
  et~al.}{2005}]{Leismann:2005rz}
Leismann T.,  Ant{\'o}n L.,  Aloy M.~A.,  M{\"u}ller E.,  Mart\'{\i} J.,
  Miralles J.~A.,    Ib{\'a}{\~n}ez J.~M.,  2005, \aap, 436, 503

\bibitem[\protect\citeauthoryear{Meliani, Keppens \& Giacomazzo}{Meliani
  et~al.}{2008}]{Meliani:2008p2303}
Meliani Z.,  Keppens R.,    Giacomazzo B.,  2008, \aap, 491, 321

\bibitem[\protect\citeauthoryear{Mignone, Plewa \& Bodo}{Mignone
  et~al.}{2005}]{Mignone:2005p2379}
Mignone A.,  Plewa T.,    Bodo G.,  2005, ApJS, 160, 199

\bibitem[\protect\citeauthoryear{Mimica, Aloy, Agudo, Mart{\'\i}, G{\'o}mez \&
  Miralles}{Mimica et~al.}{2009}]{Mimica:2009p1445}
Mimica P.,  Aloy M.-A.,  Agudo I.,  Mart{\'\i} J.~M.,  G{\'o}mez J.~L.,
  Miralles J.~A.,  2009, \apj, 696, 1142

\bibitem[\protect\citeauthoryear{Mimica, Aloy \& M{\"u}ller}{Mimica
  et~al.}{2007}]{Mimica:2007db}
Mimica P.,  Aloy M.~A.,    M{\"u}ller E.,  2007, \aap, 466, 93

\bibitem[\protect\citeauthoryear{Mimica, Aloy, M{\"u}ller \& Brinkmann}{Mimica
  et~al.}{2004}]{Mimica:2004fy}
Mimica P.,  Aloy M.~A.,  M{\"u}ller E.,    Brinkmann W.,  2004, \aap, 418, 947

\bibitem[\protect\citeauthoryear{Mimica, Aloy, M{\"u}ller \& Brinkmann}{Mimica
  et~al.}{2005}]{Mimica:2005sp}
Mimica P.,  Aloy M.~A.,  M{\"u}ller E.,    Brinkmann W.,  2005, \aap, 441, 103

\bibitem[\protect\citeauthoryear{Mimica, Giannios \& Aloy}{Mimica
  et~al.}{2009}]{Mimica:2009qa}
Mimica P.,  Giannios D.,    Aloy M.~A.,  2009, \aap, 494, 879

\bibitem[\protect\citeauthoryear{Mizuta \& Aloy}{Mizuta \&
  Aloy}{2009}]{Mizuta:2009p1213}
Mizuta A.,  Aloy M.~A.,  2009, \apj, 699, 1261

\bibitem[\protect\citeauthoryear{Mizuta, Yamasaki, Nagataki \&
  Mineshige}{Mizuta et~al.}{2006}]{Mizuta:2006p1126}
Mizuta A.,  Yamasaki T.,  Nagataki S.,    Mineshige S.,  2006, \apj, 651, 960

\bibitem[\protect\citeauthoryear{Morsony, Lazzati \& Begelman}{Morsony
  et~al.}{2007}]{Morsony:2007p1208}
Morsony B.~J.,  Lazzati D.,    Begelman M.~C.,  2007, \apj, 665, 569

\bibitem[\protect\citeauthoryear{Nakar \& Piran}{Nakar \&
  Piran}{2002}]{Nakar:2002p1323}
Nakar E.,  Piran T.,  2002, \apj, 572, L139

\bibitem[\protect\citeauthoryear{Panaitescu, Spada \&
  M{\'e}sz{\'a}ros}{Panaitescu et~al.}{1999}]{Panaitescu:1999p1022}
Panaitescu A.,  Spada M.,    M{\'e}sz{\'a}ros P.,  1999, \apj, 522, L105

\bibitem[\protect\citeauthoryear{Piran}{Piran}{2005}]{Piran:2005p2632}
Piran T.,  2005, Reviews of Modern Physics, 76, 1143

\bibitem[\protect\citeauthoryear{Rees \& Gunn}{Rees \&
  Gunn}{1974}]{Rees:1974yq}
Rees M.~J.,  Gunn J.~E.,  1974, \mnras, 167, 1

\bibitem[\protect\citeauthoryear{Rees \& Meszaros}{Rees \&
  Meszaros}{1994}]{Rees:1994ca}
Rees M.~J.,  Meszaros P.,  1994, \apjl, 430, L93

\bibitem[\protect\citeauthoryear{Rezzolla \& Zanotti}{Rezzolla \&
  Zanotti}{2001}]{Rezzolla:2001ys}
Rezzolla L.,  Zanotti O.,  2001, Journal of Fluid Mechanics, 449, 395

\bibitem[\protect\citeauthoryear{Romero, Mart\'{\i}, Pons, Ib{\'a}{\~n}ez \&
  Miralles}{Romero et~al.}{2005}]{Romero:2005zr}
Romero R.,  Mart\'{\i} J.,  Pons J.~A.,  Ib{\'a}{\~n}ez J.~M.,    Miralles
  J.~A.,  2005, Journal of Fluid Mechanics, 544, 323

\bibitem[\protect\citeauthoryear{Sari \& Piran}{Sari \&
  Piran}{1995}]{Sari:1995oq}
Sari R.,  Piran T.,  1995, \apjl, 455, L143

\bibitem[\protect\citeauthoryear{Sari \& Piran}{Sari \&
  Piran}{1997}]{Sari:1997p766}
Sari R.,  Piran T.,  1997, \apj, 485, 270

\bibitem[\protect\citeauthoryear{Sironi \& Spitkovsky}{Sironi \&
  Spitkovsky}{2009}]{Sironi:2009p414}
Sironi L.,  Spitkovsky A.,  2009, \apj, 698, 1523

\bibitem[\protect\citeauthoryear{Spada, Ghisellini, Lazzati \& Celotti}{Spada
  et~al.}{2001}]{Spada:2001p815}
Spada M.,  Ghisellini G.,  Lazzati D.,    Celotti A.,  2001, \mnras, 325, 1559

\bibitem[\protect\citeauthoryear{Tanihata, Takahashi, Kataoka \&
  Madejski}{Tanihata et~al.}{2003}]{Tanihata:2003p1291}
Tanihata C.,  Takahashi T.,  Kataoka J.,    Madejski G.~M.,  2003, \apj, 584,
  153

\bibitem[\protect\citeauthoryear{Thompson}{Thompson}{1994}]{Thompson:1994p492}
Thompson C.,  1994, \mnras, 270, 480

\bibitem[\protect\citeauthoryear{Usov}{Usov}{1992}]{Usov:1992hp}
Usov V.~V.,  1992, \nat, 357, 472

\bibitem[\protect\citeauthoryear{van~der Holst, Keppens \& Meliani}{van~der
  Holst et~al.}{2008}]{vanderHolst:2008p2553}
van~der Holst B.,  Keppens R.,    Meliani Z.,  2008, Computer Physics
  Communications, 179, 617

\bibitem[\protect\citeauthoryear{Zhang \& M{\'e}sz{\'a}ros}{Zhang \&
  M{\'e}sz{\'a}ros}{2004}]{Zhang:2004p1381}
Zhang B.,  M{\'e}sz{\'a}ros P.,  2004, International Journal of Modern Physics
  A, 19, 2385

\bibitem[\protect\citeauthoryear{Zhang, Woosley \& Heger}{Zhang
  et~al.}{2004}]{Zhang:2004p1195}
Zhang W.,  Woosley S.~E.,    Heger A.,  2004, \apj, 608, 365

\bibitem[\protect\citeauthoryear{Zhang, Woosley \& MacFadyen}{Zhang
  et~al.}{2003}]{Zhang:2003p1193}
Zhang W.,  Woosley S.~E.,    MacFadyen A.~I.,  2003, \apj, 586, 356

\end{thebibliography}

\end{document}